# A continuous wave 10 W cryogenic fiber amplifier at 1015 nm and frequency quadrupling to 254 nm

**R. Steinborn**∗, **A. Koglbauer, P. Bachor, T. Diehl, D. Kolbe**◇**, M. Stappel, and J. Walz**

*Institut für Physik, Johannes Gutenberg-Universität and Helmholtz-Institut Mainz, 55099 Mainz, Germany*
◇*Present address: Institut für Technische Physik, DLR, 70569 Stuttgart, Germany*
∗*r.steinborn@uni-mainz.de*

**Abstract:** A stable, continuous wave, single frequency fiber amplifier system at 1015 nm with 10 W output power is presented. It is based on a large mode double clad fiber cooled to liquid nitrogen temperature. The amplified light is frequency quadrupled to 254 nm and used for spectroscopy of the $6^1S \to 6^3P$ transition in mercury.

## 1. Introduction

There is a varity of interesting applications for high power laser radiation in the wavelength range from 1010 nm to 1025 nm: It can be used for laser cooling of Ytterbium doped solids [1,2] by means of optical refrigeration. The frequency quadrupled wavelength around 253.7 nm corresponds to the important $6^1S \rightarrow 6^3P$ transition in mercury, relevant for cooling and trapping of neutral mercury atoms [3, 4]. In our group, we use laser light at 253.7 nm as one of the fundamental wavelengths for four-wave sum frequency mixing in mercury vapor, to generate continuous wave vacuum-UV radiation around 121 nm for future laser cooling of trapped antihydrogen [5] as well as Rydberg excitation of trapped Ca$^+$ ions [6]. Several laser systems for generating radiation at 253.7 nm have been described, based on e.g. sum frequency mixing [7] or frequency quadrupling of semiconductor lasers [8]. Our previous setup [9–11] used a frequency quadrupled Ytterbium YAG disk laser [12]. For this system an alternative is developed, which is tunable and does not suffer from degradation of laser disks. In this paper we present a stable, high power, single frequency, continuous wave laser source at 1014.8 nm and outline the potential for a high power laser source at 253.7 nm via frequency quadrupling.

Fiber lasers and amplifiers have proven themselves as reliable systems with excellent beam quality and high output power [13–16]. Amplification in Ytterbium doped fibers is generally possible from 976 nm up to 1200 nm [17–19], but below 1030 nm amplification becomes more challenging, since the absorption cross section increases towards shorter wavelength [20]. This absorption can be reduced by cooling the fiber to low temperatures [21, 22].

Figure 1 (a) shows the absorption level scheme for Ytterbium. The splitting of the $^2F_{7/2}$ and $^2F_{5/2}$ states is caused by the Stark effekt in the host material [23]. The zero phonon line is marked with A and corresponds to the narrow peak at 976 nm in the absorption spectrum (Fig. 1 (b)). Due to the high absorption cross section this wavelength is used for pumping. Absorption from the lowest Stark-level of the $^2F_{7/2}$-state to the higher sub-levels of the $^2F_{5/2}$-state (marked with B) corresponds to the wide absorption range around 910 nm in Fig. 1 (b) and can also be used for pumping [20]. Excitations from higher sub-levels of the ground state (marked with C) are responsible for the absorption tail at longer wavelength and cause problems through signal reabsorption in fiber amplifiers and lasers working below 1030 nm. This thermal population and thus absorption in the wavelength range above 1000 nm can be significantly reduced by cooling the fiber to low temperatures. Fig. 1 (b) shows absorption spectra of a low doped Yt-terbium double clad fiber at room temperature (RT) and at liquid nitrogen temperature (LNT). For this measurement white light is coupled in the cladding of the doped fiber and the cladding absorption is observed with an optical spectrum analyzer. The reduction of the absorption in the wavelength range from 990 nm to 1050 nm at LNT is clearly visible, a significant reduction of 0.3 dB at 1015 nm can be achieved. Furthermore the zero phonon line absorption is increased at lower temperatures.

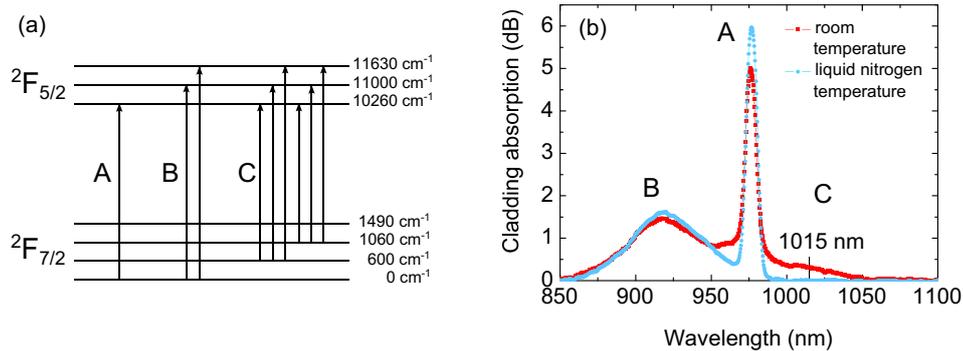

Fig. 1. (a) Level scheme of the relevant energy levels in Ytterbium. A, B and C mark the absorption transitions which can be identified in the spectrum (b). (b) Measured absorption spectrum of a 10 m low doped double clad Ytterbium fiber (Nufern LMA-YDF-10/400) at room and liquid nitrogen temperature.

## 2. Fiber amplifier at 1015 nm

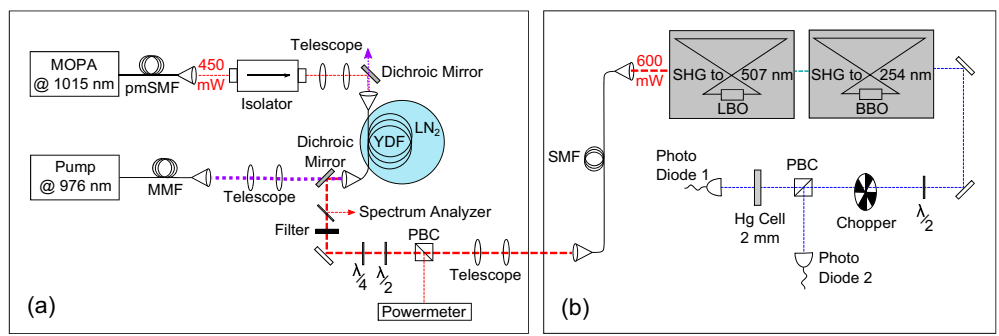

Fig. 2. Experimental setup. (a) 450 mW laser light, provided by a fiber coupled MOPA system at 1015 nm, is amplified in a cryogenic fiber amplifier up to 10 W. The doped fiber is placed in liquid nitrogen and end-pumped by a fiber coupled diode laser. The amplified light is ASE filtered and polarization analyzed. (b) A part of the amplified light is frequency doubled twice to 254 nm. For a mercury spectroscopy the beam is split at a beamsplitter cube and one part is guided through a 2 mm mercury cell to the signal photo diode, while the other part is monitored with a reference photo diode. pmSMF: polarization maintaining single mode fiber, MMF: multi mode fiber, YDF: Ytterbium doped fiber, $LN_2$: liquid nitrogen, $\lambda/2$: half-wave plate, $\lambda/4$: quarter-wave plate, PBC: polarizing beamsplitter cube, SHG: second harmonic generation, LBO: lithium triborate, BBO: $\beta$-barium borate.

The setup of the fiber amplifier is shown in Fig. 2 (a). As seed laser we use a master oscillator power amplifier setup (MOPA) consisting of a grating stabilized external cavity diode laser

(ECDL) with an optical output power of 25 mW at 1015 nm, which is pre-amplified with a tapered amplifier (TA) up to 1000 mW. The laser diode can be stabilized to an external cavity which ensures a linewidth of about 100 kHz. A polarization maintaining single mode fiber (pmSMF) guides the light to the ytterbium fiber amplifier where up to 450 mW seed power is available.

A Faraday isolator protects the end facet of the pmSMF from peak powers of the amplifier in backward direction. A dichroic mirror separates the unabsorbed pump light from the seed light, which is coupled into the core of a 7 m long large mode area ytterbium doped double clad fiber (Nufern-LMA-YDF-25/400-M, cladding absorption: ∼2 dB/m at 976 nm). The large core diameter of 25 $\mu$m of the fiber prevents stimulated Brillouin scattering. The fiber amplifier is cladding pumped from the back end with a Bragg grating stabilized fiber coupled diode laser (DILAS Compact 120/400) at 976 nm. The bare doped fiber is coiled to a diameter of approximately 15 cm and placed in a small dewar (∼3 l volume), filled with liquid nitrogen. For optical access, a length of ∼10 cm at each end of the fiber is outside the dewar and hence at room temperature.

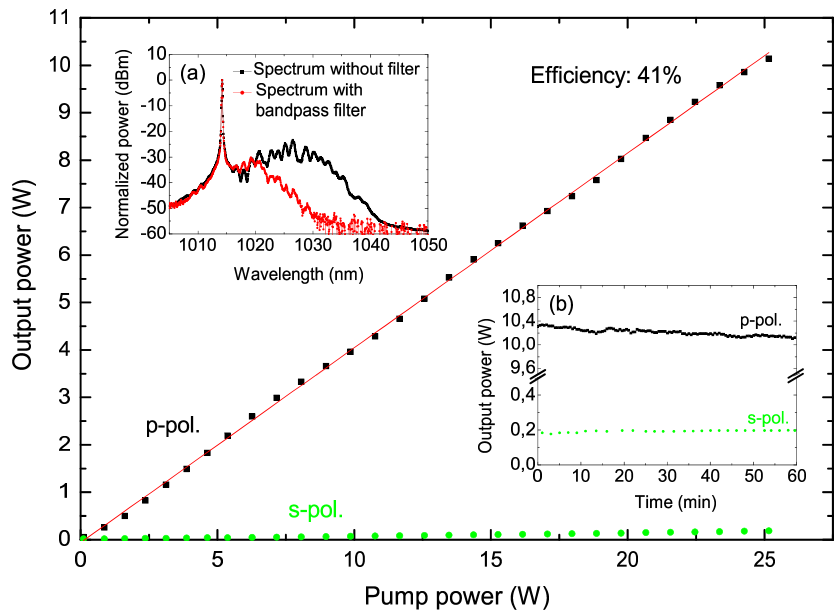

Fig. 3. P- and s-polarized output power vs. pump power. The output power is ASE and pump light filtered. The slope efficiency is 41%. Inset (a) shows the ASE spectrum with and without ASE filtering at maximum pump power. Inset (b) shows the longtime stability at maximum output power.

The output power as a function of the incident pump power is shown in Fig. 3. The maximum output power exceeds 10 W. The output power is measured after a bandpass filter to suppress amplified spontaneous emission (ASE) and reflected pump light and polarization analyzed via a polarizing beamsplitter cube. The inset (a) of Fig. 3 shows the ASE-spectrum before and after the filter, monitored with an optical spectrum analyzer, at the maximum pump power of 25 W. The ASE suppression of the unfiltered signal is better than 20 dB, the bandpass filter attenuates the ASE-signal to a suppression of 30 dB. The pump power was limited to 25 W to avoid a further increase of ASE.

Due to the large core of the doped fiber, polarization stability and single mode operation strongly depend on temperature fluctuations and bending of the fiber. For stable operation the dewar is carefully covered with a lid without any mechanical stress on the fiber and the fiber length at room temperature is reduced as far as possible. Following these precautions the output power and polarization are stable for hours (see inset (b) of Fig. 3). The fraction of non-polarizable output power is typically below 5 %. Operation time of several hours is possible without a refill of liquid nitrogen and a careful refill is possible during operation without a change in polarization, thus operation time is not limited.

The system was tested with different seed powers down to 25 mW without any problems. The results are shown in Fig. 4. For lower seed powers there is a moderate decrease in the slope efficiency, but the maximum output for 25 mW of seed power is still above 8.5 W. For decreasing seed power ASE slightly increases with no evidence of a rollover, so the seed power for the amplifier system appears to be noncritical.

Despite the large core of the doped fiber, single mode operation is easy to achieve. Signal and pump launch are stable for months and except for the position of the fiber in the liquid nitrogen, readjustment of this setup is usually not necessary.

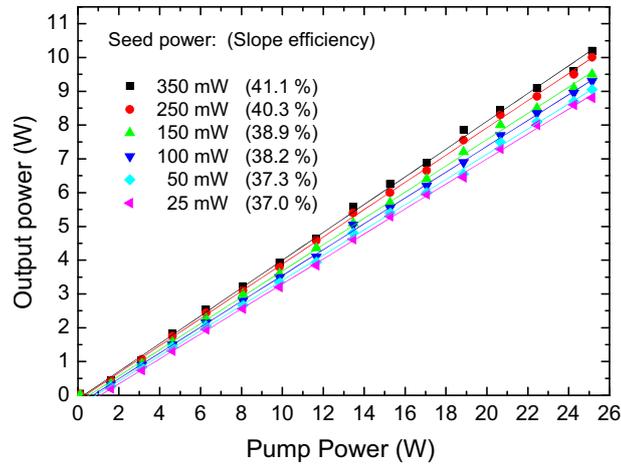

Fig. 4. Polarized output power for different seed powers as a function of the incident pump power.

## 3. Spectroscopy of the $6^1S_0 \to 6^3P_1$ transition in mercury

In order to verify the applicability of the system to nonlinear optical experiments, the amplified laser light was frequency quadrupled to 253.7 nm and we performed an absorption spectroscopy of the deep ultraviolet $6^1S_0 \to 6^3P_1$ transition in mercury.

The setup is depicted in Fig. 2 (b). To avoid stimulated Brillouin scattering (SBS), only 600 mW of the infrared light is coupled into a 20 m standard single mode fiber and guided to the frequency quadrupling system. The latter consists of two subsequent intra cavity second harmonic generations (SHG) and is described in detail in [12]. The first SHG to 508 nm uses a (type I) temperature phase matched lithium triborate crystal (LBO) as nonlinear medium in a bow tie cavity. The conversion to 253.7 nm is likewise performed in a bow tie cavity, using $\beta$-barium borate (BBO) and (type I) critical phase matching.

For background reduction from ambient light the UV laser is modulated by a chopper wheel for lock-in detection. About 1 mW of UV power is passing through a 2 mm quartz cell, filled with mercury vapor at room temperature. The distribution of isotopes corresponds to their natu-ral abundance. The transmission through the cell and a fraction of the incident light for intensity normalization are detected by identical photodiodes. Frequency scanning of the seed diode laser is performed via the feedback grating angle of the ECDL setup.

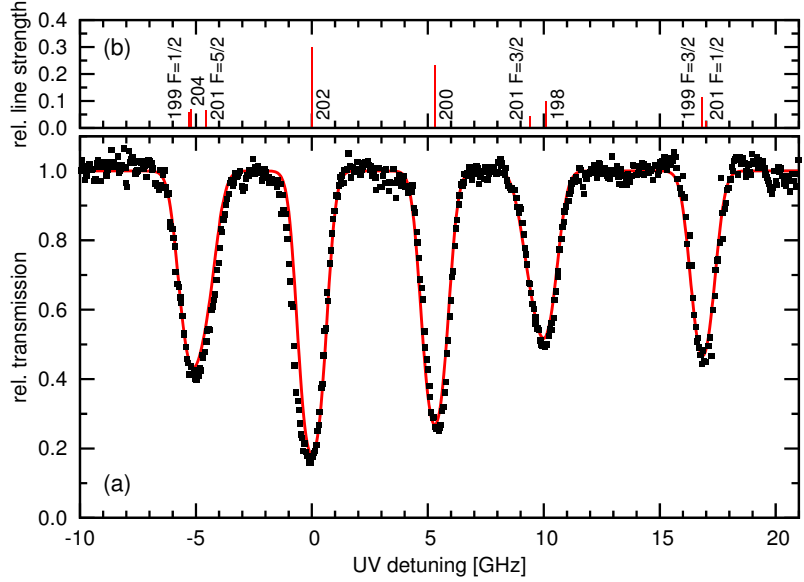

Fig. 5. (a) Measured (black squares) and theoretical (red solid line) absorption spectrum of the $6^1S_0 \to 6^3P_1$ transition in mercury vapor for a 2 mm cell at room temperature. (b) Relative strength of the mercury isotopes and their hyperfine components.

The relative transmission as a function of the UV detuning to the $6^1S_0 \to 6^3P_1$ resonance in $^{202}$Hg is shown in Fig. 5 (a) (black squares). Five absorption dips due to the mercury isotopes are clearly visible. The relative abundance and detunings of the isotopes and their hyperfine components are indicated by the red straight lines in Fig. 5 (b) above [24,25]. The red solid line beneath the data is not a fit, but a calculation of the transmission spectrum. It is based on the known transmission strengths and positions as well as the Doppler and pressure broadening for the given experimental conditions. It reproduces the data nicely, concerning linewidth, position and absorption depth.

While the graph is a combination and average of several scans, the diode laser is in principle capable of covering the whole spectral range to address all isotopes without mode hopping. Compared to the scanning rate of 15 MHz/s in the UV presented in [12], the spectrum in Fig. 5 (a) was measured with 200 MHz/s.

## 4. Conclusion

We have demonstrated a reliable, high power infrared laser source. Amplification at 1015 nm with an Ytterbium fiber amplifier was feasible by cooling the doped fiber to liquid nitrogen temperature. Despite the large temperature gradient along the fiber the amplifier is stable in polarization and output power for long time periods. The setup benefits from the convenient tuneability of the master ECDL compared to the former disk laser. In a mercury spectroscopy setup, the system has proven itself as suitable for frequency conversion to 253.7 nm. High output power of the amplifier can be achieved even with low seed power and this high output power offers the possibility of efficient second harmonic generation even in a periodic poled single pass setup. The present maximum output power of 10 W is ASE-limited and might even be enhanced in a two stage amplifier setup.

## 5. Acknowledgments

The authors acknowledge funding through the Bundesministerium für Bildung und Forschung and the ERA-NET CHIST-ERA (R-ION consortium).